\documentstyle[sprocl,epsf]{article}

\def\Journal#1#2#3#4{{#1} {\bf #2}, #3 (#4)}

\def\NPB{{\em Nucl.~Phys.~}B}
\def\PLB{{\em Phys.~Lett.~}B}
\def\PRL{\em Phys.~Rev.~Lett.~}
\def\PRD{{\em Phys.~Rev~}D}

\def\ibid{{\em ibid.~}}

\def\ra{\rightarrow}

\def\be{\begin{equation}}
\def\ee{\end{equation}}
\def\bea{\begin{eqnarray}}
\def\eea{\end{eqnarray}}

\newcommand{\nn}{\nonumber}
\newcommand{\beq}{\begin{equation}}
\newcommand{\eeq}{\end{equation}}
\newcommand{\beqa}{\begin{eqnarray}}
\newcommand{\eeqa}{\end{eqnarray}}
\def\nBR#1{{\bigl( #1 \bigr)}}
\def\ave#1{{\langle #1 \rangle}}
\newcommand\sss{\scriptscriptstyle}

\begin{document}

\title{SOURCES FOR ELECTROWEAK BARYOGENESIS}

\author{K.\ KAINULAINEN}

\address{NORDITA, Blegdamsvej 17, DK-2100, Copenhagen \O, Denmark 
         \\E-mail: kainulai@nordita.dk} 


\maketitle\abstracts{I review a computation of the baryon asymmetry 
arising from a first order electroweak phase transition in the Minimal
Supersymmetric Standard model by classical force mechanism (CFM). I focus 
on CP violation provided by the charginos and show that it is the usually 
neglected sum of the two Higgsino fields, $H_1+H_2$, which gives a larger 
contribution to the baryon asymmetry than does the combination $H_1-H_2$. 
In fact, the latter contribution is exactly zero in CFM, because it is 
associated with a phase transformation of the fields. Baryogenesis is 
found to be most effective in MSSM CFM when only $\tilde t_R$ is light, 
which lends independent support for the ``light stop scenario", and it 
remains viable for CP-violating phases as small as $\delta_\mu \sim 
{\it few} \times 10^{-3}$. }

\section{Introduction}

Although CP violation and the phase transition are known to be too weak 
for baryogenesis within the Standard Model, these problems can be overcome 
in the Minimal Supersymmetric Standard Model (MSSM). In a small region of
MSSM parameter space, corresponding to so called ``light stop" 
scenario\cite{quiros}, the transition may be strong enough to avoid the wash-out 
of baryon number by sphaleron interactions in the broken 
phase\cite{quiros,fstenf}. The sphaleron wash-out computations, 
while mired with problems associated with the infrared 
sector of gauge theories, are simple in the sense that one is dealing 
with equilibrium physics. Situation is markedly different for the theory of 
baryon production. In this case CP-violating currents are generated inside
the bubble walls, diffuse into the plasma in the unbroken phase, and bias 
sphalerons to produce the baryon asymmetry. By the very axioms of 
baryogenesis this is an inherently out-of-equilibrium system. 
As of to date, no theory exists that could 
tackle the problem in its full extent, while many scenarios have been put 
forward in an attempt to extract the leading effect in one or the other 
limit\cite{HS,JPT,Riotto,Nuria}. (However, for an ongoing project with 
the aim to self-consistently derive the transport equations for 
baryogenesis see ref.\cite{JKP}.)

Common to all methods is reducing the problem to a set of diffusion equations 
for the particle species that bias sphalerons. These coupled equations, it
is universally agreed, have the general form
\beq
	D_i \mu_i'' + v_w \mu_i' + \Gamma_i(\mu_i + \mu_j + \cdots) = S_i\;,
\label{deqn}
\eeq
where $i$ labels the particle species, $\mu_i$ is its chemical potential,
primes denote spatial derivatives in the direction ($z$) perpendicular to
the wall, $v_w$ is the wall velocity, $\Gamma_i$ is the rate of an
interaction that converts species $i$ into other kinds of particles, and
$S_i$ is the source term associated with the current generated at the
bubble wall. The essential point, and the one where little agreement exists
between different aproaches, is how to properly derive the source terms 
$S_i$ appearing in (\ref{deqn}).

In MSSM, potentially the most dominant source arises from the chargino sector. 
The CP violating effects are due to the complex parmeters $m_2$ and $\mu$ in 
the chargino mass term,
\beq
\label{chmass}
	\bar\psi_R M_\chi \psi_L = (\overline{\widetilde w^{^+}},\
	\overline{\widetilde h^{^+}}_{2} )_{R}
	\left(\begin{array}{cc} m_2 & g H_2 \\
	g H_1 & \mu 	\end{array}\right)\left(\begin{array}{c}
	\widetilde w^{^+} \\ \widetilde h^{^+}_{1} \end{array}
	\right)_{\!\!L} \,.
\eeq
Spatially varying Higgs fields cause the phase of the effective mass 
eigenstates vary nontrivially over the bubble wall. In all methods that
address the thick wall limit\cite{HS,JPT,Riotto,JKP,CJK}, one computes the 
current effected by these spatialy varying phases to leading order in an 
expansion in derivatives of the Higgs fields. 

There was an important discrepancy in the literature concerning the
derivative expansion of the chargino source.  References\cite{HN} and
\cite{cqrwv} obtained a source for the $H_1-H_2$ combination of higgsino
currents of the form
\beq
   S_{H_1-H_2} \sim {\rm Im}(m_2\,\mu)\,(H_1 H_2' - H_2 H_1'),
\label{msign}
\eeq
whereas ref.\cite{CJK}, albeit unknowingly, found the other orthogonal 
linear combination, $H_1 + H_2$, for which the result is
\beq
   S_{H_1+H_2} \sim {\rm Im}(m_2\,\mu)\,(H_1 H_2' + H_2 H_1'),
\label{psign}
\eeq
We have recently understood\cite{CK} that this disagreement about the 
sign is spurious and that all three methods actually agree with eq.\ 
(\ref{psign}); it simply was not computed by the other authors of the 
references\cite{HN,cqrwv,Riotto}.

The reason that the combination $H_1+H_2$ was not considered by the other 
authors is because it tends to be suppressed by Yukawa interactions 
and helicity-flipping interactions from the $\mu$ term in the chargino mass
matrix. Indeed, if all the interactions arising from the Lagrangian 
\beqa
        V = y \mu\tilde h_1 \tilde h_2 &+& h_2\bar u_R q_L + y \bar u_R
	\tilde h_{2L}\tilde q_L
        + y \tilde u^*_R \tilde h_{2L} q_L\nonumber\\
         &-& y \mu h_1 \tilde q^*_L \tilde u_R + yA_t\tilde q_L h_2\tilde
	u^*_R
        +\hbox{h.c.},
\label{Vint}
\eeqa
are considered to be in thermal equilibrium, they give rise to the constraints
$\xi_{H_1} - \xi_{Q_3} + \xi_T = 0 $ and $\xi_{H_2} + \xi_{Q_3} - \xi_T = 0$, 
which would damp out the effect of the source $S_{H_1+H_2}$. The rates 
$\Gamma_A$ 
of the processes coming from (\ref{Vint}) are finite however, so the equilibrium
relations are satisfied only up to corrections of order $(D_i\Gamma_A)^{-1/2}$, 
where $D_i$ is the diffusion coefficient for Higgs particles or quarks. Using 
the Higgs diffusion constant $D_h \sim 20/T$ and the Yukawa rate $\Gamma \sim 
3y^2 T/16\pi$\cite{CJK}, one finds only a mild suppression factor 
$(D_h\Gamma)^{-1/2} \sim 1$.  The source $S_{H_1-H_2}$ on the other hand 
suffers from a serious suppression: baryon number generated is (obviously) 
proportional to a spatial variation of $H_2/H_1$, but relative deviations from 
constancy of this ratio have been found to be very small, in the range 
$10^{-2}-10^{-3}$. \cite{MQS,CM}  Therefore the source $S_{H_1-H_2}$ should be 
expected to be subdominant to $S_{H_1+H_2}$ even in the models of refs.
\cite{cqrwv,MQS}. In the CFM the situation is even worse, because there the 
source for $S_{H_1-H_2}$ actually vanishes, as we shall see below.

\section{Semiclassical Boltzman equation}
\label{SCBeqn}

The classical force baryogenesis rests on particularily appealing intuitive 
picture. One assumes that the plasma in the condensate region can be 
described by a collection  of semiclassical WKB-states, following world 
lines set by their WKB-dispersion relations and corresponding canonical
equations of motion. One can then immediately write down a semiclassical 
Boltzman equation for the transport
\beq
  (\partial_t + {\bf v}_g \cdot\partial_{\bf x}  +
  {\bf F} \cdot\partial_{\bf p}) f_i = C[f_i,f_j,...].
\label{SCBE}
\eeq
where the group velocity and the classical force are given by
\beq
{\bf v}_g \equiv \partial_{{\bf p}_c} \omega  \qquad 
{\bf F} = \dot {\bf p} = \omega \dot {\bf v}_g,
\label{vgandF}
\eeq
where ${\bf p}_c$ is the canonical and ${\bf p} \equiv \omega {\bf v}_g$ 
is the physical, kinetic momentum along the WKB-world line.
Because of CP-violating effects particles and antipartices experience different
force in the wall region, $F_{\rm ap} \neq F_{\rm p}$, which leads to separation
of chiral currents.  What remains is to compute the disperson relation to obtain 
the group velocity and the force, after which the diffusion equations follow 
from (\ref{SCBE}) in a standard way by truncated moment expansion\cite{CJK}.

\subsection{Dispersion relation}

I will first consider the example of a single Dirac fermion with a spatially 
varying complex mass:
\beq
  (i\gamma^\mu\partial_\mu - m P_R - m^* P_L)\psi = 0;
  \qquad m = |m(z)| e^{i\theta(z)},
\label{dirac1}
\eeq
where $P_{L,R} = (1 \mp \gamma_5)/2$. Assuming planar walls I will also 
boost to the frame in which the momentum parallel to the wall is zero, 
$p_x=p_y=0$  (I am ignoring the effects of thermal background here). 
In this simple case it is fairly easy to solve the whole wave function 
to the first nontrivial order in the gradients,
\beq
\psi_s = \frac{|m|}{\sqrt{2p^+_s(\omega + sp_0)}}
         \left( \begin{array}{c}
             1 \\
             \frac{\omega + sp^+_s}{|m|} 
                \end{array} \right) \chi_s \; 
            e^{i\int \tilde p_s + i\frac{\theta}{2}\gamma_5 +i\phi_G},
\label{wfunc}
\eeq
where $p_0 \equiv \sqrt{\omega^2+m^2}$, $\tilde p_s \equiv p_0 + s \omega 
\theta'/(2p_0)$, $p_s^+ \equiv \tilde p_s + \omega \theta'/2$, with $\theta'
\equiv \partial_z \theta$, and 
$\sigma_3 \chi_s = s \chi_s$. The phase of the wave function in (\ref{wfunc}) 
can be written as an integral over the local (canonical) momentum:
\beq
p_c = p_0 + \frac{s \theta'}{2p_0}(\omega \pm sp_0) + \alpha_G'.
\label{DR}
\eeq
This is, of course, just the usual WKB-dispersion relation which has been 
derived in many places\cite{JPT,CJK}. The presence of an arbitrary function  
$\alpha_G'$\footnote{It may be introduced at any point by a local phase 
transition $\psi \ra e^{i\alpha_G(x)}\psi$, which leaves the lagrangian 
invariant.} 
shows explicitly, as one should expect, that $p_c$ is a gauge dependent 
quantity.  The physical quantities are gauge independent, however. For
example, in the computation of the group velocity, the gauge dependent parts 
(including the chiral rotation proportional to $\pm \theta'$) vanish because 
they are $\omega$-independent:
\beq
  v_g = \partial_{p_c} \omega = (\partial_\omega p_c)^{-1}   
      = \frac{p_0}{\omega} \nBR{1 + \frac{s m^2\theta'}{2p_0^2\omega}}
\label{vgr}  
\eeq
Similar equation holds for antiparticles, but with $\theta \ra -\theta$.
The gauge independency of the current $j^\mu = \bar \psi \gamma^\mu \psi$ 
is obvious from (\ref{wfunc}). Moreover, it is easy to show by direct 
substitution that 
\beqa
j^\mu = (1/v_g \, ; \hat {\bf p}).
\eeqa
Thus, in the absence of collisions, the WKB-particles merely follow their 
trajectories (corresponding to the stationary phase of the wave) and if they 
slow down at some point, the outcome is an increase of local density. The crux 
of the CFM is that where particles slow down, antiparticles speed up in 
relation, leading to a local particle-antiparticle bias.
    
\subsection{Physical force}
  
We still need to see how the classical force arises from the dispersion 
relation. Physically, one expects that force simply corresponds to 
acceleration, as was assumed above in Eq.\ (\ref{vgandF}). It is 
instructive to see that this force is consistent with the canonical 
equations of motion. First note that the physical momentum $p \equiv 
\omega v_g$, may be written in terms of canonical momentum as
\beq
p \; \simeq \;  p_c^\pm - \alpha^\pm - \frac{s \theta' p}{2 \omega }.
\label{connect}
\eeq
where $\alpha^\pm = \alpha_G' \pm \theta'/2$. Force acting on this
momentum is then
\beq
 F = \dot p = \dot p_c - \dot z \partial_z (\alpha^\pm + 
                           \frac{s \theta' p_k}{2 \omega }).
\label{pkeqn}
\eeq
Using the canonical equations $\dot z = v_g$  and $\dot p_c = -(\partial_z 
\omega)_{p_c}$, along with the energy conservation, one finds that
\beqa
 F = - \frac{mm'}{\omega} + \frac{s(m^2\theta')'}{2\omega^2}
   =  \omega v_g \partial_z v_g = \omega \dot v_g,
\label{force}
\eeqa
in accordance with (\ref{vgandF}). Note that while the {\em canonical 
force} $F_c \equiv -(\partial_z \omega)_{p_c}$ is obviously gauge 
dependent, the gauge parts cancel in the expression for the physical
force $F$.   Again, for antiparticles $\theta \ra -\theta$, so that
the second term in (\ref{force}) is the CP-violating force, which leads
to baryon production.

\section{Baryogensis from chargino transport}

The WKB-analysis of the chargino sector proceeds very similarly to the 
above simple example. Naturally there are some complications due to the 
additional $2\times 2$ flavour mixing structure. After a little algebra 
one finds the dispersion relation
\beqa
 p_{{\sss H}_{i_\pm}} \; = \; p_{0_\pm } 
   &\mp& 
    \frac{s(\omega + sp_{0_\pm } )}{2p_{0_\pm }}
    \frac{\Im (m_2\mu )}{m^2_\pm \Lambda } (u_1u_2' + u_2u_1') 
\nn \\
   &\mp& s_{{\sss H}_i}
    \frac{2 \Im (m_2\mu )}{\Lambda + \Delta}
                             (u_1u_2'-u_2u_1') + i \alpha_{i\pm }'
\label{DRcharg}
\eeqa
where $u_i \equiv gH_i$,  $\Lambda = m^2_+ - m^2_-$ and $\Delta =
|m_2|^2 - |\mu |^2 + u_2^2 - u_1^2$. If $m_2 > \mu$, ($m_2 < \mu$)
then the larger (smaller) mass eigenstate $m_+$ ($m_-$) corresponds to 
higgsinos.  Although promisingly $s_{{\sss H}_1} = - s_{{\sss H}_2} = 1$,
the $(u_1u_2'-u_2u_1')$-term does not source the combination $H_1-H_2$, 
because it vanishes when differentiated with respect to $\omega$. 
(It could also be absorbed into the arbitrary phase functions 
$\alpha_{i\pm}$ arising from freedom to perform field redefinitions.) 
Apart from this ``gauge'' phase, both 
higgsinos have identical dispersion relations and hence have identical 
sources in their diffusion equations, from which it follows that 
$S_{H_1-H_2} = 0$ in CFM. The nonvanishing source has a very simple 
form\cite{CK}
\beq
S_{H_1+H_2} = -\frac{s}{2}\frac{v_w D_h}{\ave{p^2/\omega^2}_\pm}
         \ave{p_z/\omega^3}_\pm
         \nBR{m_\pm^2 \theta_{\rm e}'}'',
\eeq
where $\ave{\cdots}$ refers to thermal average and $m_\pm^2 
\theta_{\rm e}' \equiv \Im (m_2\mu )(u_1u_2' + u_2u_1')/\Lambda$.
The appropriate diffusion equations have been set up and solved in 
reference\cite{CK}. The final baryon number can be written as a 
one-dimensional integral over the source
\beq
\eta_B  \propto  \frac{\Gamma_{\rm sph}}{v_w} \;
        C_{sq} \; \int_{-\infty}^\infty
       dz S_{H_1+H_2}(z) {\cal G}(z),
\label{bnumber}
\eeq
where $\Gamma_{\rm sph}$ is the Chern-Simons number diffusion rate in the 
symmetric phase\cite{Moore}, $v_w$ is the wall velocity and ${\cal G}(y)$ 
is a Greens function which I do not write explicitly here\cite{CK}. The 
parameter $C_{sq}$ encodes the essential squark spectrum dependence of our 
results: if only $\tilde t_R$ is light then $C_{sq}=5/23$. If, in addition, 
$\tilde t_L$ and $\tilde b_L$ are light then $C_{sq}=1/41$ and finally, if 
$\tilde t_L$, $\tilde b_L$ and $\tilde b_R$, and any number of other squarks 
are light then $C_{sq}=0$. This trend lends striking and entirely independent 
support for the wash-out motivated light stop scenario\cite{quiros}. 
\begin{figure}[t]
\centering
\vspace*{-4mm}
\hspace*{-3mm}
\leavevmode\epsfysize=3.0in \epsfbox{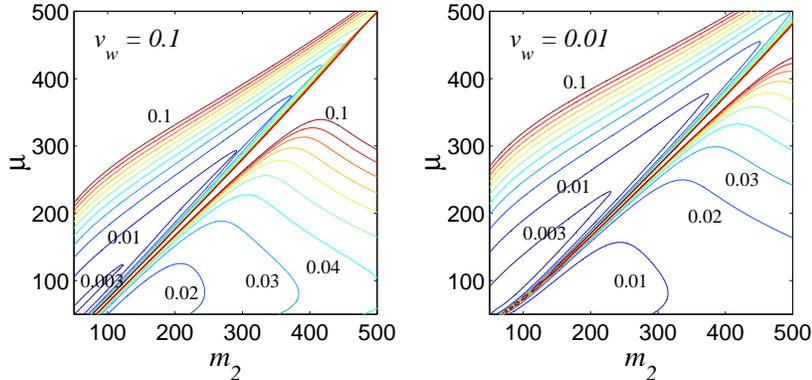}\\[-14mm]
\caption[fig1]{\label{fig1} 
The contours of $\delta_\mu$ which give rise to baryon asymmetry 
of $\eta_B = 3\times 10^{-10}$.}
\end{figure}
In Fig.\ 1 shown are the contours of $\delta_\mu = arg(\mu)$ corresponding 
to the eventual baryon to photon ratio of $\eta_B = 3\times 10^{-10}$ for 
$v_w=0.1$ and $v_w = 0.01$. Baryogenesis is seen to remain viable in the 
MSSM at least for $\delta_\mu$ as small as ${\it few} \times 10^{-3}$.

\section{Conclusions}

I have reviewed  baryogenesis via the classical force mechanism (CFM) from 
the chargino transport in the Minimal Supersymmetric Standard Model. It 
was shown that the physical quantities entering the CFM computation are 
unambiguos and independent of phase transformations on fields. It was 
pointed out that the dominant source for baryogenesis in the thick wall 
limit is the one corresponding to the linear combination of higgsinos 
$H_1 + H_2$, despite the suppression by top-Yukawa strength interactions, 
because the corresponding suppression is much milder than the suppression 
on $H_1 - H_2$ arising due to need for non-constancy of $H_2/H_1$ 
over the bubble wall\cite{cqrwv,MQS}. I suggest that this linear 
combination should lead to dominant effect also in the thin wall 
limit\cite{Nuria}.  It was also observed that CFM is most efficient for 
the case when as few squarks as possible are light, which lends 
support for the so called "light stop scenario"\cite{quiros}, necessary 
for avoiding the baryon wash-out in the broken phase.  It was finally 
shown that the CFM may be able to produce the observed baryon 
asymmetry  with the explicit CP-violating phase $\delta_\mu$ well
below present observational limits.

\end{document}